\documentstyle{mn}

\newif\ifAMStwofonts

\newcommand{\etal}{{et al.}~}

\newcommand{\pa}{\partial}
\newcommand{\f}{\frac}

\newcommand{\bfx}{\bmath{x}}

\newcommand{\bfs}{\bmath{s}}

\newcommand{\bfv}{\bmath{v}}
\newcommand{\bfV}{\bmath{V}}

\newcommand{\bfu}{\bmath{u}}
\newcommand{\vnabla}{\bmath{\nabla}}

\newcommand{\eps}{{\epsilon}}
\newcommand{\bc}{\begin{center}}
\newcommand{\be}{\begin{equation}}
\newcommand{\ee}{\end{equation}}
\newcommand{\ec}{\end{center}}

\def\dd{\, {\mathrm d}}
\def\vx{\bfx}
\def\vs{\bfs}
\def\vv{\bfv}
\def\vu{\bfu}
\def\vlhat{\hat{\bmath x}_3}
\def\vxhat{\hat{\bfx}}
\def\vshat{\hat{\bfs}}

\title{On the vorticity of  flow in redshift space}

\author[Chodorowski \& Nusser]
      {Micha{\l} J.\ Chodorowski$^1$ and Adi Nusser$^2$
\\ 
$^1$Copernicus Astronomical Center,
      Bartycka 18, 00--716 Warsaw, Poland\\
$^2$Physics Department, The Technion-Israel Institute of Technology }
\begin{document}
\maketitle
\begin{abstract}
Given an irrotational (vorticity free) velocity field in real space,
we prove that, in the distant observer limit and in the absence of
multi-valued zones, the associated velocity field in redshift space is
also irrotational. The proof does not rely on any approximation to
gravitational dynamics. The result can be particularly useful for the
analysis of redshift distortions and for reconstruction methods of
cosmological velocity fields from galaxy redshift surveys, in the
nonlinear regime. Although the proof is restricted to the distant
observer limit, we show that the {\sc potent} method can be modified to
derive the full {\em real} space velocity field as a {\em function} of
the {\em redshift space coordinate}, thus avoiding spatial Malmquist
biases.
\end{abstract}


\section{Introduction}
\label{sec:intro}
Peculiar motions (deviations from pure Hubble flow) could cause
significant deviations between the distribution of galaxies in
redshift space (s-space) and real space (x-space) (e.g. Kaiser 1987).
Because of these deviations (redshift distortions), the known linear
and nonlinear relations between dynamical fields (e.g. peculiar
velocity and density) in x-space cannot directly be applied in
s-space.  Most analysis of redshift distortions has been done using
linear theory (Kaiser 1987, Hamilton 1992, Nusser \& Davis 1994,
Fisher et. al. 1995, Heavens \& Taylor 1995).  Although various
approximations to dynamics in x-space have been developed, little
progress has been made in extending these approximations to s-space
(but see Hivon \etal 1995, Fisher \& Nusser 1996, Taylor \& Hamilton
1996, Hatton \& Cole 1998, Scoccimarro et al. 1999). The common
belief that the velocity field in s-space is not
irrotational\footnote{Vorticity is the curl of a vector field.  An
irrotational vector field is a field with zero vorticity at any point
in space. This is equivalent to having zero circulation around an
arbitrary closed path or to being derivable from a scalar potential
function.}, even in the distant observer limit (DOL), has been the
main hindrance in the development of nonlinear analysis methods in
s-space.

According to Kelvin's circulation theorem (e.g., Landau \& Lifshitz
1959), if the initial velocity field is irrotational, i.e., curl-free,
then, in the absence of shell-crossing, the nonlinear velocity field
is also irrotational.  Kelvin's theorem concerns velocity fields in
x-space.  This means that the x-space velocity field, $\bfv(\bfx)$, at
any point $\bfx$ can be derived from a potential, $\Phi_v(\bfx)$,
i.e.,
\be
\bfv(\bfx)=-\vnabla_x \Phi_v(\bfx) \,.
\label{eq:potent}
\ee 
This relation has been the corner stone in the majority of large scale
structure studies. It is the basis of the {\sc potent} method (cf. Dekel
1994 for a review) and the modal expansion method (Nusser \& Davis
1995) for recovering velocity fields from raw data of radial peculiar
velocities. It is also the basis of various nonlinear approximations
to dynamics (Bernardeau 1992, Gramann 1993, Mancinelli \& Yahil 1995,
Chodorowski 1997, Chodorowski \& {\L}okas 1997, Chodorowski et
al. 1998, Bernardeau et al. 1999). Here we will show that in the DOL
and in the absence of multi-values zones (regions with overlapping
streams in s-space) the s-space velocity field associated with an
irrotational flow in x-space, is curl-free.  Although the proof is
valid only in the DOL, the limit in which most studies of redshift
distortions are done, we show how the {\sc potent} method can be generalized
to provide x-space velocity potentials as a function of the {\em
s-space\/} coordinate and thereby avoiding Malmquist bias corrections.

In section 2 we prove our claim by showing that the circulation is
zero and also by direct calculation of the curl of the s-space
velocity field.  In section 3 we present a relation between the
s-space and x-space velocity potentials and show how {\sc potent} can be
modified to provide x-space velocity and potential fields as a
function of the s-space coordinate.  In section 4 we summarize the
results and briefly discuss possible applications. 
\enlargethispage*{1000pt}

\section{The proof}
Let $\vv(\vx)$ be the {\em x-space\/} comoving peculiar velocity as a
function \pagebreak 
of the x-space comoving coordinate $\vx$, both expressed 
in $\mathrm km\; s^{-1}$. The s-space comoving coordinate, $\vs$, is then 
\begin{equation}
\vs=\vx + \left [\vxhat \cdot \vv(\vx) \right] \vxhat \,,
\label{redshift:general}
\end{equation}
where $\vxhat=\bfx/x $ is a unit vector in the line of sight direction.
Denote by $\vu(\vs)$ the {\em s-space\/} peculiar velocity  
as a function of 
$\vs$. Consider a patch of matter at an x-space position 
$\vx$ moving with velocity $\vv(\vx)$. The s-space position of this
patch of matter is given by (\ref{redshift:general}) and its 
s-space velocity is
\begin{equation}
\vu(\vs)= \vv(\vx) .
\label{vel:id}
\end{equation}
In the DOL, we restrict the analysis to a region of space where the
separation between any two points is small compared to the distance of
the observer from the region. Therefore  the lines of
sight to all points in the region can be approximated to have the
same direction, which we arbitrarily choose to be the unit vector
$\vlhat$ in a Cartesian coordinate system $(x_1,x_2,x_3)$.  Therefore,
substituting $\vxhat=\vlhat$ in equation (\ref{redshift:general}), we
find
\begin{equation}
\vs=\vx + v_3(\vx) \vlhat ,
\label{dol}
\end{equation}
where $v_3=\vv \cdot \vlhat$ is the  component of $\vv$  
along the line of sight.

\subsection{The circulation: $\oint \vu \cdot \dd \vs =0$}
If $\vv$ is irrotational then its circulation 
around an arbitrary closed path in $\vx$ space is zero.
So we have,
\begin{equation}
\oint \vv(\vx) \cdot \dd \vx =0 .
\label{circx}
\end{equation}

We will show  that, in the DOL, the velocity 
field $\vu(\vs)$ is also irrotational.
Consider the circulation 
\begin{equation}
C=\oint \vu(\vs) \cdot \dd \vs .
\end{equation}
In the absence of multi-valued zones, we can use
 (\ref{vel:id}) and (\ref{dol}) to express
 the integral in this equation in terms of $\vv$ and $\vx$,
\begin{equation}
C=\oint \vv(\vx) \cdot \dd \left[\vx +v_3(\vx) \vlhat \right] ,
\label{circs}
\end{equation}
where the integration is over a closed path in x-space, obtained 
by applying the transformation (\ref{dol}) on every point on the 
closed  path in s-space.
Because of (\ref{circx}) the relation (\ref{circs}) reduces to
\begin{equation}
C=\oint \vv(\vx) \cdot \vlhat \dd v_3(\vx) =\oint v_3 \dd v_3 =0 .
\end{equation}
Therefore the circulation around an arbitrarily chosen path in s-space
is zero.  According to Stokes theorem this implies that curl of
$\vu$, or its vorticity, is zero.  Nevertheless, in the next
subsection, we show, by direct calculation, that if
$\vnabla_{\bfx} \times  \vv(\bfx)=0$ 
then
$\vnabla_{\bfs} \times  \bfu(\bfs)=0$.
 
\subsection{The curl: ${\vnabla}_{\bfs}\times \bfu(\bfs) =0$}

The vorticity of the s-space velocity field is 
\be
\bfV=\vnabla_{\bfs}\times \bfu(\bfs) \,.
\ee
In the absence of multi-values zones the mapping between $\vs$ and
$\vx$ is one-to-one and therefore the relation (\ref{vel:id}) and 
the chain rule can be used to yield
\be
\bfV = 
\vnabla_{\bfs} \times \bfv[\bfx(\bfs)] = 
\f{\pa \bfx}{\pa \bfs}\vnabla_{\bfx} \times \bfv(\bfx)
\,.
\ee
Hence, in a Cartesian coordinate system, the vorticity's $i$-th
component is
\be
V_i = 
\eps^{ijk} \f{\pa x_l}{\pa s_j} \f{\pa}{\pa x_l} v_k 
\,,
\label{eq:vort_cart}
\ee
where $\eps^{ijk}$ is Levy-Civita anti-symmetric symbol. 
{}From equation~(\ref{dol}) we have
\be
\f{\pa s_i}{\pa x_j} = \delta_{ij} + \delta_{i3} v_{3,j}
\,,
\label{eq:ds_dr}
\ee
where $\delta_{ij} $ is the Kronecker delta function and 
$f_{,j} \equiv (\pa/\pa x_j) f$. To compute the vorticity we
need to know the inverse matrix to $\pa \bfs/\pa \bfx$. Its
computation is straightforward and the result is

\be
\f{\pa x_j}{\pa s_k} = \delta_{jk} - J^{-1} \delta_{j3} v_{3,k}
\,,
\label{eq:dr_ds}
\ee
where $J = 1 + v_{3,3}$ is the {\em exact\/} Jacobian of the
transformation from real to redshift space in the DOL. 

Using equation~(\ref{eq:dr_ds}) in~(\ref{eq:vort_cart}) yields
\begin{eqnarray}
V_i 
&=& \eps^{ijk} \left( \delta_{lj} - J^{-1} \delta_{l3} v_{3,j} \right)
\f{\pa}{\pa x_l} v_k \nonumber \\
&=& \eps^{ijk} 
\left(\f{\pa}{\pa x_j} - J^{-1} v_{3,j} \f{\pa}{\pa x_3} \right)
v_k \nonumber \\
&=& \eps^{ijk} v_{k,j} - J^{-1} \eps^{ijk} v_{3,j} v_{k,3} 
\,.
\label{eq:vort_alg}
\end{eqnarray}
The x-space velocity field is irrotational, so the first term on the
RHS of the last line vanishes. For the same reason, $v_{3,j} =
v_{j,3}$; therefore, being a contraction of the anti-symmetric symbol
with a symmetric tensor, the second term also vanishes.  Therefore,
$\bfV=0$.

\section{Relation between s-space and x-space velocity potentials}

The s-space velocity potential $\Phi_u(\vs)$ is defined by
\begin{equation}
\vu(\vs)=-\vnabla_{\bfs} \Phi_u(\vs) \,,
\label{potentials}
\end{equation}
and can be written, up to an additive constant, as 
\begin{equation}
\Phi_u(\vs)=-\int^{\vs} \vu( {\tilde {\vs}}) \cdot \dd  \tilde {\vs} \,,
\label{phiu:s}
\end{equation}
where the integration is along any path connecting $\vs$ to 
an arbitrary  fixed reference point. 
Using (\ref{vel:id}) and (\ref{dol}), the last equation 
yields
\begin{equation}
\Phi_u(\vs)=\Phi_v[\bfx(\bfs)] -\frac{1}{2}v^2_3[\bfx(\bfs)] \,,
\label{phiu:x}
\end{equation}
in which $v_3[\bfx(\bfs)] $ can be replaced by $u_3(\bfs)$.

Our proof of potential flow is valid only in the DOL. Suppose however
that we would like to obtain the x-space velocity potential from all
sky measurements of radial peculiar velocities of galaxies in our
cosmological neighborhood.  The {\sc potent} machinery works in x-space. It
first smoothes the measured velocities in x-space, then it corrects
for spatial (homogeneous and inhomogeneous) Malmquist biases and
integrates the smooth radial velocities along the radial direction to
obtain the x-space potential field.  We now show how to obtain the
x-space velocity potential from the measured radial velocities as a
function of the galaxies redshifts. This is worthwhile as the x-space
velocity field, $\vv$, presented as function of the s-space coordinate
does not suffer from the spatial Malmquist biases (e.g., Schechter
1980, Tully 1988).

Working with  spherical coordinates, $\bfx=(x,\theta,\phi)$,
we   write (Bertschinger \& Dekel 1989)
\begin{equation}
 \Phi_v(x,\theta,\phi) =
-\int_0^{x}v_{\mathrm rad}(\tilde x,\theta,\phi) \dd \tilde x \,,
\end{equation}
where $v_{\mathrm rad}$ is the smoothed radial velocity field and 
the integration is along the radial direction.  
Using (\ref{vel:id}) and the general relation (\ref{redshift:general})
between x-space and s-space coordinates the last equation
becomes,
\begin{equation}
 \Phi_v[s(x),\theta,\phi] =
-\int_0^{s(x)}u_{\mathrm rad}(\tilde s,\theta,\phi) \dd [\tilde s-u_{\mathrm rad}(\tilde s)] \,,
\label{phiv:rad}
\end{equation}
where $s(x)=x+v_{\mathrm rad}$ and $u_{\mathrm rad}$ is the radial
component of the s-space velocity $\bfu$. In arriving at
(\ref{phiv:rad}) we have assumed that velocities are measured relative
to the Local Group motion and so $s(x=0)=0$.  Therefore the x-space
potential as a function of the s-space coordinate can be expressed in
terms of the s-space radial velocities as 
\be
\Phi_v(\bfs)=-\int_0^{s}u_{\mathrm rad}(\tilde s,\theta,\phi) \dd
\tilde s + \frac{1}{2}u^2_{\mathrm rad}(\bfs) \,.  
\ee 
The full x-space velocity field in
terms of the s-space coordinate can be derived from the x-space velocity 
potential in the following way 
\be
\bfv=-\vnabla_{\bfx} \Phi_v(\bfs)=-\frac{\pa \bfs}{\pa \bfx}
\vnabla_{\bfs} \Phi_v(\bfs) \,, 
\ee 
where $\pa \bfs /\pa \bfx$ is computed by rewriting
(\ref{redshift:general}) in the form $\bfx=\bfs-u_{\mathrm rad}
(\bfs)\vshat$.

\section{summary and discussion}

We have shown that the s-space velocity field in the distant observer
limit and in the absence of shell crossing is irrotational. The result
can have important implications on methods of nonlinear reconstruction
of velocity from density and vice versa in redshift space. It is also
relevant to estimating cosmological parameters from the apparent
anisotropies of clustering in redshift space (cf. Hamilton 1998, for a
review). For example, Nusser \& Davis (1994) have presented a
relation, based on the Zel'dovich approximation (Zel'dovich 1970),
between the velocity and density in s-space. By expressing the s-space
velocity in terms of a potential, this relation can now be solved
iteratively to obtain the velocity field associated with a given
redshift galaxy distribution. Furthermore, our result can simplify
calculations of redshift space dynamical relations based on
perturbation theory (e.g., Chodorowski 1999).

The distant observer limit is achieved by restricting the analysis to
far away regions with sufficiently small opening angle.  Future
redshift surveys such as the Anglo-Australian 2dF and the Sloan
Digital Sky Survey (Gunn \& Knapp 1993) will be sufficiently deep that
the distant observer limit can be easily achieved for large fractions
of the surveys.

Finally, we have shown that the {\sc potent} method can be modified to
derive the full x-space velocity field as a function of the s-space
coordinate, thus avoiding spatial Malmquist biases. This result is
general, i.e. its validity is not restricted to the limit of the
distant observer.

\section*{Acknowledgments}
MC thanks Rom\'an Scoccimarro for stimulating discussions. MC also
acknowledges partial support by the Polish State Committee for
Scientific Research grants No.~2.P03D.008.13 and 2.P03D.004.13.


\begin{thebibliography}{}
\bibitem{} Bernardeau F., 1992, ApJ, 390, L61
\bibitem{} Bernardeau F., Chodorowski M. J., {\L}okas E. L., Stompor R., 
  Kudlicki A., 1999, preprint astro-ph/9901057 
\bibitem{} Bertschinger E., Dekel A., 1989, ApJ, 336, L5
\bibitem{} Chodorowski M. J., 1997, MNRAS, 292, 695 
\bibitem{} Chodorowski M. J., 1999, MNRAS, 308, 640 
\bibitem{} Chodorowski M. J., {\L}okas E. L., 1997, MNRAS, 287, 591
\bibitem{} Chodorowski M. J., {\L}okas E. L., Pollo A., Nusser A., 
    1998, MNRAS, 300, 1027 
\bibitem{} Dekel A., 1994, ARA\&A,32, 371
\bibitem{} Fisher K.B., Lahav O., Hoffman Y., Lynden-Bell D., Zaroubi S.,
1995, MNRAS, 272, 885
\bibitem{} Fisher K.B., Nusser A., 1996, MNRAS, 279, L1
\bibitem{} Gramann M., 1993, ApJ, 405, L37
\bibitem{} Gunn J.E., Knapp G.,1993, in {\em Sky Surveys}, Soifer B.T. (ed),
Astronomical Society of the Pacific Conference Series \# 43, p267
\bibitem{} Hamilton A.J.S., 1992, ApJ, 385, L5
\bibitem{} Hamilton A.J.S., 1998, {\em Proceedings Ringberg Workshop
on Large-Scale Structure}, Hamilton D., (ed), Kluwer, Dordrecht 
\bibitem{} Hatton S., Cole S., 1998, MNRAS, 296, 10
\bibitem{} Heavens A.F., Taylor A.N., 1995,MNRAS, 275, 483
\bibitem{} Hivon E., Bouchet F. R., Colombi S., Juszkiewicz R., 
	1995, A\&A, 298, 643
\bibitem{} Kaiser N., 1987, MNRAS,227,1
\bibitem{} Landau L.D., Lifshitz E.M., 1959, {\em Fluid Mechanics},
Pergamon Press
\bibitem{} Mancinelli P. J., Yahil A., 1995, ApJ, 452, 75 
\bibitem{} Nusser A., Davis M., 1994, ApJ, 421, L1
\bibitem{} Nusser A., Davis M., 1995, MNRAS, 276, 1391
\bibitem{} Taylor A.N., Hamilton A.J.S., 1996, MNRAS, 282, 767
\bibitem{} Schechter P.L., 1980, AJ, 85, 801
\bibitem{} Scoccimarro R., Couchman H. M. P., Frieman J. A., 1999,
	ApJ, 517, 531 
\bibitem{} Tully R.B., 1988, Nature, 334, 209
\bibitem{} Zel'dovich Ya.B., 1970, A\&A, 5, 20
\end{thebibliography}
\end{document}